# Accurate biomolecular simulations account for electronic polarization

Josef Melcr [1*] and Jean-Philip Piquemal [2,3,4*]

[1] Groningen Biomolecular Sciences and Biotechnology Institute and The Zernike Institute for Advanced Materials, University of Groningen, 9747 AG Groningen, The Netherlands

[2] Laboratoire de Chimie Théorique, Sorbonne Université, UMR7616 CNRS, Paris, France.

[3] Institut Universitaire de France, 75005, Paris, France.

[4] Department of Biomedical Engineering, the University of Texas at Austin, Austin, TX 78712, USA.

Contacts: j.melcr@rug.nl (Groningen); jean-philip.piquemal@sorbonne-université.fr (Sorbonne)

## Abstract

In this perspective, we discuss where and how accounting for electronic many-body polarization affects the accuracy of classical molecular dynamics simulations of biomolecules. While the effects of electronic polarization are highly pronounced for molecules with an opposite total charge, they are also non-negligible for interactions with overall neutral molecules. For instance, neglecting these effects in important biomolecules like amino acids and phospholipids affects the structure of proteins and membranes having a large impact on interpreting experimental data as well as building coarse grained models. With the combined advances in theory, algorithms and computational power it is currently realistic to perform simulations with explicit polarizable dipoles on systems with relevant sizes and complexity. Alternatively, the effects of electronic polarization can also be included at zero additional computational cost compared to standard fixed-charge force fields using the electronic continuum correction, as was recently demonstrated for several classes of biomolecules.

In molecular dynamics simulations, the interactions between molecules are described with approximate potentials known as force fields that mimic the true Born-Oppenheimer energy hypersurface. Among these methods, pairwise additive potentials are very popular for modeling biomolecules such as proteins, lipids or nucleic acids.(Lopes et al., 2015; Ponder and Case, 2003) The current standard force fields(Huang and MacKerell, 2013; Maier et al., 2015; Robertson et al., 2015), however, neglect important physical many-body effects such as the electronic polarization, charge transfer or many-body dispersion (cited in decaying magnitude order).(Kleshchonok and Tkatchenko, 2018) Although such models have provided valuable insight into many phenomena from various fields including biology, chemistry, biophysics or material sciences, there are several important cases in which accounting for polarizability is crucial.

## Pitfalls of non-polarizable force fields

The limited predictive accuracy of non-polarizable force fields led the molecular modeling community to develop new generation "polarizable" force fields (Antila et al., 2019; Gresh et al., 2007; Jing et al., 2019; Jorgensen, 2007; Kleshchonok and Tkatchenko, 2018;

Martinek et al., 2018; Melcr et al., 2018, [CSL STYLE ERROR: reference with no printed form.]; Piquemal and Jordan, 2017; Shi et al., 2015; Stone, 2013) able to include the missing physics with a special focus on the polarizability effects. Although such techniques are now widely used in fields studying highly charged ionic liquids (Bedrov et al., 2019), their application cannot be limited only to such extreme cases. For instance, neglecting the effects of the electronic polarizability in important biomolecules like amino acids, nucleic acids and phospholipids affects the structure of proteins(Duboué-Dijon et al., 2018a; Jiao et al., 2008; Shi et al., 2009), DNA(Babin et al., 2006) and membranes(Catte et al., 2016; Harder et al., 2009; Melcr et al., 2018) having a large impact on interpreting experimental data (Berkowitz et al., 2012; Böckmann and Grubmüller, 2004; Eisenberg et al., 1979; Feigenson, 1986; Hauser et al., 1977; Javanainen et al., 2017; Kurland et al., 1979; Lund et al., 2008; Magarkar et al., 2017; Mattai et al., 1989; Melcrová et al., 2016; Roux and Bloom, 1990, 1991; Vacha et al., 2009) as well as building coarse grained models. Importantly, these results show that the electronic polarization yields non-negligible effects also at overall neutral molecules(Gresh et al., 2007; Melcr et al., 2018).

Secondary structure of proteins is to a large extent determined by an intricate network of hydrogen bonds. The description of hydrogen bonds in standard force fields, however, does not contain important contributions, e.g. from polarization and partially covalent character (Babin et al., 2006). It was demonstrated in many cases including structure of water (Dang, 1998), binding of ligands to proteins(Friesner, 2005; Jiao et al., 2008) and protein folding and unfolding(Célerse et al.; Freddolino et al., 2010; Huang and MacKerell, 2014; Lemkul et al., 2016; Morozov et al., 2006; Piana et al., 2011, 2014) that polarizability contributes significantly to the accuracy of simulations of structures with hydrogen bonds. Also, salt bridging between amino acids is likely overestimated in strength when the effects of polarization are not included (Ahmed et al., 2018; Célerse et al., 2019; Debiec et al., 2014; Friesner, 2005; Mason et al., 2019; Vazdar et al., 2013). For instance, the interaction of acidic side chains of glutamate and aspartate with cations is overestimated in strength in classical non-polarizable force fields(Duboué-Dijon et al., 2018a; Patel et al., 2009), while treatment of polarizability in solvent relaxation affects salt bridge dissociation(Célerse et al., 2019). Taken together, the secondary and tertiary structural arrangements in the simulations of proteins are likely biased to certain preferred configurations due to the lack of polarizability depending on the chosen parametrization strategy. (Freddolino et al., 2009, 2010; Piana et al., 2011, 2014)

Membrane proteins form a large part of cellular proteome and are in direct contact with amphiphilic cellular membranes, which influence their structure and activity.(Lee, 2004) Membranes themselves are crucial cell organelles which define the inner resp. outer cellular environment. They are predominantly composed of amphiphilic phospholipids, which self-assemble into stable bilayer structures.(Harayama and Riezman, 2018) The force fields for phospholipids have been tuned to the level that the simulations of commonly used simplified model lipid membranes can reproduce a large variety of experimentally measured properties, phenomena and structural features including lipid self-diffusion, x-ray scattering patterns, bilayer thickness, area per phospholipid and acyl chain order parameters(Pluhackova et al., 2016).

This could make an impression that the currently available non-polarizable lipid force fields provide comparable accuracy to the models with explicit polarization at a fraction of the computational cost. While the non-polarizable models yield accurate results in many cases,(Chowdhary et al., 2013a; Lucas et al., 2012) simulation studies have revealed that such models gradually lose their predictive accuracy with increasing complexity beyond model systems used during their parametrization, e.g. when membranes are put into contact with buffers of physiological ionic strengths.(Catte et al., 2016) For instance, improvements in the electrostatics of phospholipid membranes have a great impact on the membrane dipole potential, permeation of water through membranes, and viscosity of organic liquids(Harder et al., 2009; Venable et al., 2019). Moreover, the interactions between phospholipids and cations, especially divalent cations like calcium, are overestimated in the classical non-polarizable models. (Antila et al., 2019; Catte et al., 2016; Melcr et al., 2018)

In general, the structure of divalent cations complexes that are widespread in biosystems is traditionally problematic in non-polarizable simulations(Kohagen et al., 2015). In contrast, simulations with explicit or implicit treatment of polarization yield comparable accuracy to DFT-based ab-initio calculations and neutron scattering experiments, as was demonstrated for biologically relevant divalent cations $Ca^{2+}$ and $Mg^{2+}$ (Martinek et al., 2018; Piquemal et al., 2006b; Wu et al., 2010). While accounting for the electronic polarization overall improves the predictive accuracy of simulations in general, it is not sufficient in some cases like zinc chloride ion pairing, where more complex physics beyond "mere" electronic polarization is at play. (Duboué-Dijon et al., 2018b; Gresh et al., 2005, 2007; Piquemal et al., 2007)

## Implicit treatment of electronic polarization via electronic continuum correction

The necessity of polarizability and screening in modeling lipid bilayers has been an issue from the very beginning of computational modeling of model membranes. The first pioneering works on phospholipid bilayers document the need of including polarizability and extra screening in the development of the first models, which was achieved at that time through an empirical scaling factor for the partial atomic charges of the phospholipids(Egberts et al., 1994). A similar strategy supported by continuum theory was used in the recent developments of phospholipid force fields, which implicitly account for the electronic polarization using Electronic continuum correction (ECC).(Leontyev and Stuchebrukhov, 2009, 2010a; Martinek et al., 2018; Mason et al., 2012; Pegado et al., 2012; Pluhařová et al., 2013) Despite the approximate treatment of the polarizability using ECC, such lipid force fields provide accurate interactions between phospholipid bilayers and cations in agreement with experiments.(Melcr et al., 2018) In particular, in the case of the neutral phosphatidylcholine (PC), ECC improved the cation binding affinity for monovalent and divalent cations reaching agreement with experiments(Melcr et al., 2018), while for negatively charged phosphatidylserine (PS) it has also improved the overall structure of the phospholipid and the interactions with other lipids.(Antila et al., 2019; Melcr et al.)

Electronic continuum correction is a very efficient alternative to otherwise computationally demanding explicit modeling of electronic polarization.(Bedrov et al., 2019) The accuracy of the ECC method was shown to yield promising results on several polar organic solvents(Lee and Park, 2011; Leontyev and Stuchebrukhov, 2010b, 2012; Vazdar et al., 2013), while it proved to be necessary yet sufficient for an accurate description of the structure of several monovalent and divalent ions in aqueous solutions(Mason et al., 2012; Pegado et al., 2012; Pluhařová et al., 2013). To date, the array of force fields utilizing ECC has grown from a wide range of biologically relevant ions (Kohagen et al., 2014, 2015; Martinek et al., 2018), to protein moieties(Duboué-Dijon et al., 2018a; Mason et al., 2019; Vazdar et al., 2013) and whole phospholipid molecules(Melcr et al., 2018) making realistic simulations of e.g. membrane proteins at physiological ionic conditions possible.

In ECC all particles are assumed to have equal polarizabilities and the electric field and electron density within each particle is homogenous.(Leontyev and Stuchebrukhov, 2009) Such approximations simplify the calculations of the polarization to such an extent that it can be simply included in the interactions as a pre-determined charge-scaling factor(Leontyev and Stuchebrukhov, 2009), which is derived from the high-frequency dielectric constant of electrons, $\varepsilon_{el}$, as $1/\sqrt{\varepsilon_{el}} \approx 0.75$ for aqueous solutions. Importantly, $\varepsilon_{el}$ is close to 2 for a wide variety of biologically relevant environments meaning that even interfaces like biological membranes do not give rise to large gradients. Despite the coarseness of the approximations, the effects of electronic polarization are described sufficiently well for a variety of biologically relevant molecules in a condensed phase.(Duboué-Dijon et al., 2017; Duboué-Dijon et al., 2018a; Martinek et al., 2018; Melcr et al., 2018) Moreover, ECC accounts for the effects of electronic polarization at zero additional computational cost compared to standard fixed-charge force fields. Although, a new generation of simulation codes performing large scale simulations with explicit polarization models starts to emerge(Lagardère et al., 2018), ECC yields the benefit of employing the widely adopted and already highly optimized codes for classical MD.

The common implementation of ECC via charge rescaling profoundly resembles an empirical scaling factor, which, obviously, reduces the interaction of charged molecules. From both the derived ECC theory(Leontyev and Stuchebrukhov, 2010a) and its applications, which compare ECC to also other methods (Martinek et al., 2018; Pegado et al., 2012), it is however clear that the improvements pertinent to ECC can be attributed to the electronic polarization. For instance, interactions of sulphate anions were directly compared between simulations with ECC, solvent shell model (Rick and Stuart, 2003) and ab-initio calculations(Pegado et al., 2012). This comparison has revealed that ECC performed comparably well to the other methods at a fraction of the computational cost. Moreover, ECC was concluded as preferable over the explicit solvent shell model for sulphate anions as it was closer to the structures from ab-initio calculations(Pegado et al., 2012).

## Capturing effects beyond electronic polarization

The accuracy of the implicit methods including ECC is limited and gradually falls short in cases, which do not adhere to the assumed approximations. For instance, the complex electronic structure of $Zn^{2+}$ makes it difficult to capture the ion pairing of zinc chloride with ECC unless specific *ad hoc* interaction terms between the ions are introduced. (Duboué-Dijon et al., 2018b) Hence, resorting to more accurate modeling strategies including explicit polarizable dipoles — or even effects beyond electronic polarization — becomes necessary in such cases.

The water structure around $Zn^{2+}$ in bulk solution and its free energy of hydration is correctly reproduced by the AMOEBA force field with explicit polarizable dipoles, but it still does not capture the fine details of zinc chloride ion pairing. The reason for that is that $Zn^{2+}$ exhibits considerably large charge transfer effects prefiguring what is happening with transition metals where back-donation effects become important. (Gresh et al., 2005, 2007; Piquemal et al., 2007) Simulations then need to utilize more complex polarizable force fields able to separately evaluate the different physical contributions. Indeed, short-range electrostatics in such systems is anything but classical as it is strongly affected by quantum penetration effects in the overlap region(Gresh et al., 2005, 2007; Piquemal et al., 2003, 2006a; Wang et al., 2015). On the contrary, many-body polarization interactions which are usually cooperative (i.e. the total energy being larger that the purely additive contributions) do not behave in such a way.(Gresh et al., 2007, 2016; Jing et al., 2018; Zhang et al., 2012) Divalent metal cations in particular locally reverse the physical trends and exhibit net anticooperativity as the total energy becomes smaller than the sum of individual contributions. For example SIBFA (Sum of Interactions Between Fragments Ab initio computed) incorporates a many-body explicit charge transfer (Gresh et al., 2005, 2007; Piquemal et al., 2007) and a penetration correction for electrostatics (Narth et al., 2016; Piquemal et al., 2003), and is able to deal with such difficult systems.

Such effects also exist with variable magnitude in biomolecular simulations, and resorting to more accurate methods employing physics even beyond explicit polarization will be likely required for predictive accuracy in many cases, e.g. metalloproteins, which shall be interesting playgrounds for such modeling.(Gresh et al., 2007, 2016; Jing et al., 2018; Zhang et al., 2012) Improvements in capturing correct physics is a general trend in current developments, and besides SIBFA, the AMOEBA force field is gradually evolving into the AMOEBA+ potential, which additionally includes such physical effects (Liu et al., 2019). Moreover, several other general polarizable potentials are emerging(Das et al., 2019; Huang et al., 2017; Rackers and Ponder, 2019) indicating the start of next-generation polarizable force fields development (Duke et al., 2014; Piquemal et al., 2006a; Piquemal and Cisneros, 2016).

## Are polarizable simulations computationally tractable?

This being said the question remains: is there any practically achievable perspective application of such advanced models to meaningfully large simulations of biologically relevant systems? — Certainly yes. If the use of polarizable models has been doomed by their computational cost for years, things have dramatically improved. In terms of computational requirements, the approaches utilizing Drude particles (Lopes et al., 2013) traditionally appeared more feasible compared to explicit point dipole approaches (Lagardère et al., 2014, 2015), as their computational cost in standard high-performance codes was higher by a factor 2 to 4 depending on implementation and reference settings compared to non-polarizable force fields (Jiang et al., 2011), while the explicit point dipoles models were roughly twice slower. However, such models cannot utilize long time-steps because of their use of extended Lagrangian, which practically imposes a speed limit(Albaugh and Head-Gordon, 2017; Wang and Skeel, 2005). In contrast, utilizing advanced algorithms for solving polarization and dynamical integration is possible within explicit point dipole approaches leading to strong speed increases to the performance level of Drude approach (even for higher-level multipolar electrostatics approaches such as AMOEBA) when compared to usual non-polarizable models simulation.(Lagardère et al., 2019) However, the numerous available point dipole force fields (AMOEBA, SIBFA etc…) had in practice another handicap besides their computational cost: they were not available in high performance/production codes such as GROMACS or NAMD (Phillips et al., 2002; Van Der Spoel et al., 2005).

This situation has gradually changed in recent years. First, in link with the improved multi-timestep integration, the key mathematical problem of solving the point dipole equations using iterative methods was alleviated using new non-iterative approaches such as the Truncated Conjugate Gradient (TCG-1) (Aviat et al., 2017a, 2017b) that allows for a fix cost evaluation of polarization. When coupled to an analytical evaluation of gradient such an approach fully preserves energy and, hence, allows for long time step simulations. Second, the availability of massively parallel MPI codes able to efficiently use supercomputers using 3D domain decomposition techniques such as Tinker-HP (Jolly et al., 2019; Lagardère et al., 2018) (the high performance engine of the Tinker molecular package(Rackers et al., 2018)) shed first rays of light at the end of the tunnel leading towards simulations of biologically relevant large systems on long enough timescales with explicit polarization. Moreover, GPU accelerated implementations of AMOEBA in OpenMM(Huang et al., 2018) and Tinker-OpenMM (Harger et al., 2017) are available whereas the support of hybrid (multi)CPUs-GPUs is coming in Tinker-HP (O. Adjoua et al., personal communication).

Overall, methodology has made a key progress and will continue in this direction for all types of polarizable force fields as the accessible computer power quickly increases reducing therefore the computational gap with additive potentials. Whereas specialized highly accurate water potentials based on many-body expansions emerge such as MBPOL (Riera et al., 2019) and allow for a better understanding of fine physical effects in clusters and bulk water, the availability of general polarizable force fields such as AMOEBA offering water (Ren and Ponder, 2003), ions, organochlorine compounds (Mu et al., 2014), proteins and nucleic acids (Shi et al., 2013; Zhang et al., 2018) now enables performing enough sampling to achieve highly accurate and biologically meaningful simulations. The Drude approaches parametrization is expanding as well (Chowdhary et al., 2013a, 2013b; Lamoureux et al., 2003; Lopes et al., 2013). Moreover, accelerated sampling methods start to be applied also to polarizable approaches (Célerse et al., 2019) offering improved simulation capabilities and access to accurate and fast evaluation of free energies of binding thanks to GPUs.(Harger et al., 2017) Such capabilities allow to tackle hard systems as in the case of the Phosphate binding mode of the Phosphate-binding protein where it was possible to highlight the critical effect of the buffer solution ending a long standing controversy thanks to free energy computations. (Qi et al., 2018)

## Summary

In summary, we have presented several important classes and case studies of biomolecules, where including polarizability is an important factor for the simulation accuracy. Cytosolic environment in cells is mostly composed of water solutions of ions, for which polarizability is necessary for the accurate description of the solvated structure of ions, their pairing and interaction with other biomolecules. (Duboué-Dijon et al., 2017; Duboué-Dijon et al., 2018a; Martinek et al., 2018; Mason et al., 2012; Melcr et al., 2018; Pegado et al., 2012; Piquemal et al., 2006b; Pluhařová et al., 2013; Wu et al., 2010) Polarizability is an important factor for accurate interactions between amino acids, namely salt bridges between them, which are overestimated in strength in current non-polarizable force fields(Ahmed et al., 2018; Célerse et al., 2019; Friesner, 2005; Mason et al., 2019; Vazdar et al., 2013). Moreover, polarizable force fields yield a better description of the hydrophobic effect and hydrogen bond networks in proteins, which to a large extent determine the dynamic structure and conformational changes of proteins(Célerse et al., 2019; Dill et al., 1995; Fitch et al., 2002; Freddolino et al., 2010; García-Moreno et al., 1997; Huang and MacKerell, 2014; Lemkul et al., 2016; Morozov et al., 2006; Piana et al., 2011, 2014; Venable et al., 2019). Polarizability is necessary for accurate structure and interactions of both neutral and charged phospholipids, which constitute a dominant part of cellular membranes.(Catte et al., 2016; Harder et al., 2009; Melcr et al., 2018)

The representation of electronic polarization in classical MD simulations can vary largely with Drude and induced point dipoles approaches on one side and continuum approximations on the other (Baker, 2015; Bedrov et al., 2019; Cieplak et al., 2009; Jing et al., 2019; Lemkul et al., 2016; Leontyev and Stuchebrukhov, 2011; Lopes et al., 2009; Schröder, 2012; Shi et al., 2015). With the advances in both computational power together with theory and algorithms it is practically achievable to perform simulations with explicit polarizable dipoles on systems with relevant sizes and complexity.(Bedrov et al., 2019; Lagardère et al., 2019; Loco et al., 2019; Qi et al., 2018)  In particular, it is currently realistic to perform simulations with explicit polarization at time scales, which are competitive to the standard fixed-charge simulations.(Célerse et al., 2019; Lagardère et al., 2018; Lemkul et al., 2016) Moreover, approximate implicit solutions like ECC, which circumvent the computational costs of explicit polarization, gradually gain on popularity and provide a promising solution for a variety of applications in biomolecular simulations.(Duboué-Dijon et al., 2017; Duboué-Dijon et al., 2018a; Martinek et al., 2018; Mason et al., 2019; Melcr et al., 2018) Finally, as fully variational polarizable embeddings are now possible in  hybrid QM/MM molecular simulations (Loco et al., 2016, 2017, 2019), one can expect that hybrid explicit polarization/ECC simulations will be possible in the near future offering a multi-level global treatment of polarization across very large complex molecular systems.

Biomolecules in the real world cannot turn off their polarizability. Hence, molecular dynamics simulations, which aim to give a realistic, robust and predictive results, cannot afford to neglect this important contribution to the electrostatic interaction. Currently, polarizable force fields for a large variety of biomolecules and simulation codes implementing polarizability exist and are readily available to solve various biophysical problems.(Bedrov et al., 2019; Célerse et al., 2019; Chowdhary et al., 2013a; Duboué-Dijon et al., 2017; Duboué-Dijon et al., 2018a; Jing et al., 2019; Lagardère et al., 2018; Lemkul et al., 2016; Liu et al., 2019; Martinek et al., 2018; Melcr et al., 2018; Wu et al., 2010; Zhang et al., 2018) We expect that the popularity of such approaches will grow and will become a common tool in biomolecular research in the near future.

**Acknowledgments**

This work has received funding from the European Research Council (ERC) under the European Union's Horizon 2020 research and innovation programme (grant agreement No 810367), project EMC2.